\newcommand{\beq}{\begin{equation}}
\newcommand{\eeq}{\end{equation}}

\newcommand{\id}
 {i\kern.06em\hbox{\raise.25ex\hbox{$/$}\kern-.60em$\partial$}}

\newcommand{\bs}{/\kern-.52em b}
\newcommand{\qs}{/\kern-.52em s}

\newcommand{\p}{\partial}
\newcommand{\yp}{^{\prime}}

\newcommand{\dd}
{\kern.06em\hbox{\raise.25ex\hbox{$/$}\kern-.60em$\partial$}}

\newcommand{\ep}{\epsilon}
\newcommand{\Tr}{\mathop{\rm Tr}\nolimits}

\documentstyle[12pt]{article}
\begin{document}
\title{Exact properties of the chemical potential-density
 relation at finite temperature in the Hubbard model
\thanks{S.S. Feng is on leave of absence from the Physics Department,
Shanghai University, 201800, Shanghai, China}
\author{{Sze-Shiang Feng$^{\P}$, Ferdinando Mancini$^{\S}$}\\
\P.{\small {\it CCAST(World Lab.), P.O. Box 8730, Beijing 100080}}\\
\P.{\small {\it Department of Astronomy and Applied Physics}}\\
          {\small {\it University of Science and Technology
    of China, 230026, Hefei, China}}\\e-mail:sshfeng@yahoo.com\\
\S.{\small {\it Universit$\grave{a}$ degli Studi di Salerno-Unit$\grave{a}$
   IFNM di  Salerno}}\\
   {\small {\it Dipartmento di Scienze Fisiche "E.R. Caianiello"
    84081 Baronissi Salerno, Italy}}\\
    email: mancini@physics.unisa.it}}
\maketitle
\baselineskip 0.3in
\begin{center}
\begin{minipage}{135mm}
\vskip 0.3in
\baselineskip 0.3in
\begin{center}{\bf Abstract}\end{center}
  {We draw some rigorous conclusions about the functional
  properties of the $\mu-\rho$ relation in the Hubbard model
   based on symmetry considerations and
  unitary transformations. It is shown that the charge susceptibility reaches
  its local extreme at half-filling. Exact expressions are obtained
  in two limiting cases.
  \\PACS number(s): 75.10.Lp,71.20.Ad,74.65.+n,74.20.Mn
   \\Key words: $\mu-\rho$ relation,Hubbard model}

\end{minipage}
\end{center}
\vskip 1in
\section{Introduction}
\indent Exact results are of great interests in discussions of
strong correlations
\cite{s1}. The important one-dimensional strongly correlated
 models such as Hubbard model and Kondo model
 can be exactly solved by means of Bethe ansatz\cite{s2}\cite{s3}.
Nevertheless, exact knowledge of higher dimensional models are
very rare. Remarkably, the $\eta$-paring mechanism\cite{s4} can be employed
to discuss the possible existence of the off-diagonal long range order(ODLRO)
\cite{s5} at both zero and finite temperatures \cite{s6}\cite{s7}.
The exact knowledge about the chemical-potential relation and magnetization
, which are important aspects of the statistics, is still called for. Because
of this situation, a number of approximate approaches
are applied, as well as computational simulations. It should be realized
that every reliable approximation must coincide with the exact one
qualitatively. That is, it should have the properties
of the exact solution. Therefore, the exact properties
of the $\mu-\rho$ relation should be
studied. This is the motivation of this paper. The layout of this
paper is as follows. Section 2 discusses the general exact
properties of $\mu-\rho$ relation. The last section is devoted to conclusional
remarkes.\\
\section{Properties of the $\mu-\rho$ relation}
\indent The Hamiltonian of the generic Hubbard model 
on lattice $\Lambda$ is
\beq
H=\sum_{ij}\sum_{\sigma}t_{ij}c^{\dag}_{i\sigma}c_{j\sigma}
+U\sum_in_{i\uparrow}n_{i\downarrow}
\eeq
and the grand canonical partition function is
\beq
Z(\mu,\beta;t,U)=\Tr\exp[-\beta(H-\mu N)]
\eeq
where $c^{\dag}_{i\sigma}$ and $c_{i\sigma}$ are the creation
and annihilation operators of the electrons with spin $\sigma=
\uparrow, \downarrow$ at site $i$. The hopping matrix
$\{t_{ij}\}$ is required
to be real and symmetric. The number operators are $n_{i\sigma}
=c^{\dag}_{i\sigma}c_{i\sigma}$, while the $U$ denotes the on-site
Coulomb interaction. It is further assumed that the the lattice
$\Lambda$
is bipartite in the sense that it
can be divided into sublattices {\bf A} and {\bf B} such that $t_{ij}=0$ whenever
$\{ij\}\in {\bf A}$ or $\{ij\}\in {\bf B}$.
$N=\sum_{i,\sigma}
n_{i\sigma}$ is the  number operator of the electrons.
Though the desired $\mu-\rho$ relation is of the form $\mu=\mu(\rho,\beta)$
in statistics, the properties to be discussed below is that of the inverse
relation $\rho=\rho(\mu,\beta)$.\\
\indent We first derive a recursion relation concerning the moments
of the number operators. For this purpose, we make use of the complete
particle-hole transformation
\beq
Pc_{i\sigma}P^{-1}=\ep(i)c^{\dag}_{i\sigma}
\eeq
where $\ep(i)=1/-1 $ if $i\in {\bf A}/{\bf B}$.
Under this transformation, the grand canonical Hamiltonian undergoes
the following transformation
\beq
P(H-\mu N)P^{-1}=H+U(N_{\Lambda}-N)-\mu(2N_{\Lambda}-N)
\eeq
where $N_{\Lambda}$ is the
total number of sites of the lattice $\Lambda$.
Therefore, the partition function satisfies the following relation
\beq
Z(\mu,\beta;t,U)=\exp[-\beta(U-2\mu)N_{\Lambda}]Z(U-\mu,\beta;t,U)
\eeq
We have accordingly for $n=1,2,3,\cdots$
\begin{eqnarray}
<N^n>_{\mu}&=&Z^{-1}\Tr PN^nP^{-1}\exp[-\beta P(H_0-\mu N)P^{-1}]\\
&=&Z^{-1}\Tr(2N_{\Lambda}-N)^n\exp[-\beta P(H_0-\mu N)P^{-1}]\\
&=&<(2N_{\Lambda}-N)^n>_{U-\mu}
\end{eqnarray}
This relation holds also locally. Consider the local density operator
$n_i=\sum_{\sigma}c^{\dag}_{i\sigma}c_{i\sigma}$, using the transformation (3), 
the same reasoning leads to
\beq
<n^{k}_i>_{\mu}=<(2-n_i)^k>_{U-\mu}
\eeq
This local relation can not be obtained from (8)
directly. In particular, we have at half-filling
for which $\mu=\mu_c=U/2$ from (9) $<n_i>_{\mu_c}=1$. The global relation
(8) gives then
\beq
<N^n>_{\mu_c}=<(2N_{\Lambda}-N)^n>_{\mu_c}
\eeq
That is
\beq
[1+(-1)^{n+1}]<N^n>_{\mu_c}=\sum^{n}_{l=1}C^l_n(2N_{\Lambda})^l(-1)^{n-l}
<N^{n-l}>_{\mu_c}
\eeq
where $C^l_n$ is the combinatorial number:$C^l_n=\frac{n!}{l!(n-l)!}$. 
The first three cases  for odd $n$ are
\begin{eqnarray}
 <N>_{\mu_c}&=&N_{\Lambda}\\
<N^3>_{\mu_c}&=&3N_{\Lambda}<N^2>_{\mu_c}-2N_{\Lambda}^3\\
<N^5>_{\mu_c}&=&5N_{\Lambda}<N^4>_{\mu_c}-20N_{\Lambda}^3<N^2>
_{\mu_c}+16N_{\Lambda}^5
\end{eqnarray}
In general, eq.(10) turn out to be identities  for even $n$ and it provides
non-trivial information only for odd $n$. On the other hand, since the
fluctuation
\beq
\frac{1}{\beta}\frac{\p}{\p\mu}<N>_{\mu}=<N^2>_{\mu}-<N>^2_{\mu}
\eeq
is of order $<N>$, we may introduce
a function $\alpha$ via
\beq
\frac{1}{\beta}\frac{\p}{\p\mu}<N>_{\mu}=\alpha(\mu)<N>_{\mu}
\eeq
One can see that the derivatives of $<N>$ with respect
to $\mu$ exist at every order. Obviously, one has
\beq
\alpha\ge0
\eeq
The function $\alpha$ is related to the compressibility which is defined
by
\beq
\kappa=-\frac{1}{V}(\frac{\p V}{\p P})_{T,N}
\eeq
as follows. Since $\rho=N/V$, we may express $\kappa$ as
\beq
\kappa=\frac{1}{\rho}(\frac{\p\rho}{\p P})_{T,N}=
\frac{1}{\rho}(\frac{\p\rho}{\p\mu})_{T,N}(\frac{\p\mu}{\p P})_{T,N}
\eeq
Using the thermodynamic relations\cite{s8}
\beq
\mu=(\frac{\p\Phi}{\p N})_{P,T}, \,\,\,\,\,\,\,
V=(\frac{\p\Phi}{\p P})_{T,N}
\eeq
where $\Phi$ is the thermodynamic potential, we have
\beq
(\frac{\p\mu}{\p P})_{T,N}=(\frac{\p V}{\p N})_{T,P}=\frac{1}{\rho}
\eeq
therefore
\beq
\kappa=\frac{1}{\rho^2}(\frac{\p\rho}{\p\mu})_T=\frac{\alpha\beta}{\rho}
\eeq
This relation is more thermodynamical than statistical. It holds for
both fermi liquids and non-fermi liquids. For the electron gas at zero
temperature, this relation can be derived in another way\cite{s9}.
In the Hubbard model, the density is defined by $\rho=<N>/N_{\Lambda}$.
Since $<N>$ has maximum value $2N_{\Lambda}$ and is a monotonically
increasing function of $\mu$, we have
\beq
\lim_{\rho\rightarrow 2}\alpha=0
\eeq
\indent We now prove some more properties of $\alpha$. Firstly, for a homogeneous
system, we have $<N>=N_{\Lambda}<n_i>$ and $<N^2>=\sum_{i,j}<n_in_j>=
N_{\Lambda}\sum_{i}<n_in_0>$. Thus (15) implies that
\beq
\kappa=\frac{\beta}{\rho^2}N_{\Lambda}[\frac{1}{N_{\Lambda}}
\sum_i<n_in_0>-<n_0>^2]
\eeq
which provides another way to evaluate the compressibility.
Define the density-density correlation function
\beq
C_{i\sigma,j\sigma^{\yp}}(\mu)=<n_{i\sigma}n_{j\sigma^{\yp}}>
-<n_{i\sigma}><n_{j\sigma^{\yp}}>
\eeq
the relation
\beq
C_{i\sigma,j\sigma^{\yp}}(\mu)=C_{i\sigma,j\sigma^{\yp}}(U-\mu)
\eeq
can be obtained (not only for a homogeneous system) by the local relation(9).
Secondly, applying the complete particle-hole transformation to eq.(9),
we have
\beq
<(2N_{\Lambda}-N)^2>_{U-\mu}-<2N_{\Lambda}-N>^2_{U-\mu}
=\alpha(\mu)<2N_{\Lambda}-N>_{U-\mu}
\eeq
Expanding this relation and using eq(15)
for $\mu\rightarrow U-\mu$, one can obtain
\beq
[\alpha(\mu)+\alpha(U-\mu)]<N>_{U-\mu}=\alpha(\mu)2N_{\Lambda}
\eeq
Thirdly, differentiate (28) with respect to $\mu$, we have
\beq
[\alpha\yp(\mu)+\alpha\yp(\nu)(-1)_{|\nu=U-\mu}]<N>_{U-\mu}
+[\alpha(\mu)+\alpha(U-\mu)]\frac{\p}{\p\nu}<N>_{\nu}(-1)_{|\nu
=U-\mu}=\alpha\yp(\mu)2N_{\Lambda}
\eeq
Therefore, at $\mu=\mu_c$
\beq
2\alpha(\mu_c)(-1)\frac{\p}{\p\mu}<N>_{|\mu_c}=\alpha\yp(\mu_c)
2N_{\Lambda}
\eeq
Hence we obtain another relation
\beq
\alpha^{\prime}(\mu_c)=-\beta\alpha^2(\mu_c)
\eeq
Using this relation, one can prove that the charge susceptibility $\chi_c:=
\frac{\p\rho}{\p\mu}$ takes its
extreme at half-filling, i,e,
\beq
\chi_c^{\yp}(\mu_c)=0
\eeq 
This is a rigorous result and is valid in all cases at finite temperature.
Unfortunately, we can not know whether it is a minimum or a maximum at the present time.
If it is a minimum and the value $\chi_c(\mu_c)=0$, the system will demonstrate a Mott
transition.\\ 
\indent We next study the Taylor expansion of $\rho=\rho(\mu)$
around $\mu=\mu_c$. From eq(8) we have
\beq
\rho(\mu)=2-\rho(U-\mu)
\eeq
If we define
\beq
f(\mu)=\rho(\mu_c+\mu)-1
\eeq
then
\beq
f(\mu)=-f(-\mu)
\eeq
i.e. $f(\mu)$ is an odd function of $\mu$. For finite temperature, $\rho=\rho(\mu)$ is an analytic function and can
be expanded in powers of $\mu-\mu_c$, thereby, we may infer that the
Taylor expansion of $\rho(\mu)$ must be of the form {\it at finite temperature}
\beq
\rho(\mu)=1+\sum^{\infty}_{k=0}
\frac{\rho^{(2k+1)}(\mu_c)}{(2k+1)!}(\mu-\mu_c)^{2k+1}
\eeq
The average values $<N^n>_{\mu_c}$ for both odd and even $n$ must
be known in order to know the Taylor coefficients in (36). Unfortunately,
the recursion relation eq(8) can not
 provide us with useful information about $<N^n>_{\mu_c}$ for even $n$
. Otherwise, the exact form of $\mu-\rho$ relation can be immediately
obtained after the sum is implemented. In any case, equation of motion
is needed in order to calculate $<N^k>$ for the moment.\\
\indent Though the general exact $\mu-\rho$ relation is not accessible so far,
the specific expressions for two limiting cases: non-interacting
case and atomic case are obtainable nevertheless.
In the first 
 case, $U=0$,  if the lattice is periodic, we can
work in the ${\bf k}$-space and we have
\beq
<n_{{\bf k}\sigma}>=\frac{1}{e^{\beta[t({\bf k})-\mu]}+1}
\eeq
where $t({\bf k})$ is the Fourier transform of $t_{ij}$. Therefore
the $\mu-\rho$ relation is
\beq
\rho=\frac{1}{N_{\Lambda}}\sum_{{\bf k}\sigma}<n_{{\bf k}\sigma}>
\eeq
In the second case, $t_{ij}=0$, the partition function can be calculated
\beq
Z(\beta,\mu;U)=\Tr\exp[-\beta(U\sum_in_{i\uparrow}n_{i\downarrow}-\mu N)]
=[1+2e^{\beta\mu}+e^{\beta(2\mu-U)}]^{N_{\Lambda}}
\eeq
Therefore, the average density is
\beq
\rho(\mu, \beta)=\frac{2e^{\beta(\mu-\mu_c)}e^{\beta\mu_c}+
2e^{2\beta(\mu-\mu_c)}}{1+2e^{\beta(\mu-\mu_c)}e^{\beta\mu_c}+
e^{2\beta(\mu-\mu_c)}}
\eeq
This relation is also exact for the case of infinite hopping range\cite{s10}.
It can be seen that 
the Taylor coefficients must be in general both $t_{ij}$ and $U$ dependent as
can be seen from the two special cases (38) and (40). The inverse relation
at finite temperature $\mu=\mu(\rho,\beta)$ can be obtained. For $\rho\not=2$
we have
\beq
\mu=\frac{1}{\beta}\ln[\rho-1+\sqrt{(\rho-1)^2+\rho(2-\rho)e^{-2\beta\mu_c}}]
-\frac{1}{\beta}\ln(2-\rho)+2\mu_c
\eeq
while for $\rho=2$, we can obtain only $\mu=\infty$ at finite temperature otherwise
the denorminator in (40) is finite and can be multiplied by both sides and
in this way we will reach the unacceptable conclusion $e^{\beta\mu}=-1$. It should be
noted that discontinuities may appear at zero temperature. For instance,
if $\mu_c>0$ we have from (40)
\[
\lim_{\beta\rightarrow\infty}\mu=\left\{
\begin{array}{ll}
\lim_{\beta\rightarrow\infty}\frac{1}{\beta}\ln[\frac{\rho(2-\rho)}{2|\rho-1|}
e^{-2\beta\mu_c}]+2\mu_c=0  &if\ \rho<1\\
\mu_c & if\ \rho=1\\
2\mu_c & if\ \rho>1
\end{array}
\right.
\]
while for $\mu_c<0 $ and $\rho\not=2, \lim_{\beta\rightarrow\infty}
\mu=\mu_c$.\\
\indent For $t_{ij}\not=0$, numerical and analytical results show that there
is a critical value $U_c$ such that for $U>U_c$ at $T=0$, there
is a gap in the density of states and a discontinuity
of $\mu$\cite{s11}.\\
\indent Introducing the dimensionless parametres $\tilde{t}=\beta t,
\tilde{\mu}=\beta\mu, \tilde{U}=\beta U, \rho$ is a function of these
parametres: $\rho=\rho(\tilde{t},\tilde{\mu},\tilde{U})$. Since under
 the Lieb-Wu transform in \cite{s2}, $t\rightarrow -t$, the dependence
 of $t$ is in fact through $t^2$. Hence we have
\beq
\rho(\beta,\mu;t,U)=\rho(\tilde{\mu};\tilde{t},\tilde{U}
)=1+\sum^{\infty}_{k=0}
\frac{\rho^{(2k+1)}(\tilde{\mu}_c;\tilde{t}^2,\tilde{U})}
{(2k+1)!}(\tilde{\mu}-\tilde{\mu}_c)^{2k+1}
\eeq

\section{Discussions}
\indent We discussed exact properties of the  chemical potential-density 
relation by means of the particle-hole transformations and the symmetries of the Hubbard model. Similar discussion applies to Kondo lattice model and
periodic Anderson models as in \cite{s7}. These exact
features can help us to evaluate the reliabilities of the various
conclusions from different approximations. Yet, these properties
are still not enough to determine the relations completely.
It can be inferred from the exact solutions in the two limiting
cases that the general exact answer must be extremely complicated.\\
\vskip 0.3in
\underline{Acknoledgement}This work is supported by the Funds for Young
Teachers of Shanghai Education Committee, the National
Science Foundation of China under Grant.No. 19805004 and by
Instituto Nazionale di Fisica Della Materia, Unit$\grave{a}$ di Salerno
. 

\end{document}